\documentclass[prl,twocolumn,showpacs,letterpaper,showpacs,superscriptaddress]{revtex4-1}
\usepackage{graphicx,amsmath,amssymb,amsfonts,latexsym,color,dcolumn,bm,epsfig,subfigure}

\usepackage{graphicx,amsmath,amssymb,amsfonts,latexsym,color,dcolumn,bm,epsfig,subfigure}
\renewcommand{\imath}[0]{\mathrm{i}}
\newcommand{\mathbfh}[1]{\hat{\mathbf{#1}}}

\begin{document}

\title{Vacuum Incalescence}

\author{Francesco Intravaia}
\email{francesco.intravaia@mbi-berlin.de}
\affiliation{Max-Born-Institut, 12489 Berlin, Germany}

\begin{abstract} 
In quantum theory the vacuum is defined as a state of minimum energy that is devoid of particles but still not completely empty. It is perhaps more surprising that its definition depends on the geometry of the system and on the trajectory of an observer through space-time. Along these lines we investigate the case of an atom flying at constant velocity near a planar surface. Using general concepts of statistical mechanics it is shown that the motion-modified interaction with the electromagnetic vacuum is formally equivalent to the 
interaction with a thermal field having an effective temperature determined by the atom's velocity and distance from the surface. This result suggests new ways to experimentally investigate the properties of the quantum vacuum in non-equilibrium systems and effects such as quantum friction.
\end{abstract}


\maketitle


The advent of the quantum theory has deeply changed our idea of empty space by obliging us to reformulate our notion of the vacuum from a state of nothingness to a state roiling with fluctuations. 
This has led to new fundamental questions but also to new interesting
predictions such as the Casimir effect \cite{Casimir48}. 
Nowadays, effects related to the properties of vacuum are no longer theoretical curiosities and have attracted growing attention from the experimental community for their multiple implications in science \cite{Decca07,Wilson11a,Steinhauer14} and technologies \cite{Intravaia13,Zou13}. 
In the 1970s another unexpected twist occurred. In their seminal papers, Fulling, Davies and Unruh 
\cite{Fulling73,Davies75,Unruh76}
predicted that a detector moving through the quantum vacuum with uniform acceleration 
$a$  effectively ``feels'' the surrounding field as if it were in a state of thermal equilibrium at
temperature 
\begin{equation}
T_{\rm HU}=\frac{\hbar a}{2\pi k_{\rm B} c}~.
\label{UnruhTemp}
\end{equation}
Here $\hbar$, $k_{\rm B}$, and $c$ denote, respectively, the Planck constant, the Boltzmann 
constant, and the speed of light in vacuum. This expression, sometimes called the Hawking-Unruh temperature, is connected to another seminal 
contribution 
from Hawking, who, a few years before, by mixing quantum electrodynamics and general relativity, 
demonstrated that thermal photons are created by the black hole's gravitational field \cite{Hawking74}. Their temperature is formally identical to \eqref{UnruhTemp}, where, however, the acceleration is in this case replaced by the surface gravity of the black hole. 
While the Casimir effect has highlighted the reliance of vacuum on geometrical boundary conditions \cite{Intravaia13,Intravaia12a},
the Fulling-Davies-Unruh (FDU) effect has played a crucial role in our understanding of 
quantum field theory by showing that the quantum vacuum also depends on the reference frame of 
the observer 
\cite{Crispino08,Boyer84,Alsing04}. 

Acceleration is an essential ingredient for the FDU effect. Indeed, the covariant form of quantum field theory implies that a motion through vacuum with uniform 
velocity is physically equivalent to the stationary case (Lorentz invariance). Consequently, despite quantum fluctuations, an object moving in empty space at constant velocity will preserve its motion undisturbed. This is, however, no longer 
valid if one can define a privileged frame of reference with respect to which 
one can determine the dynamical and kinematical properties of the system. 
For example, if the motion of an object takes place within a thermal field at finite temperature 
$T$, a force acting on the object tends to bring it to rest in the frame set by the (blackbody) 
radiation 
\cite{Mkrtchian03,Dedkov05,Volokitin15a}. 
A similar situation occurs when the object is moving at uniform velocity near another body: 
In certain circumstances Lorentz invariance is broken, giving rise to a frictional force (quantum 
friction) on the moving object even when the temperature is zero
\cite{Pendry97,Volokitin02,Dedkov02a,Scheel09,Barton10b,Zhao12,Pieplow13,Intravaia14,Intravaia15,Hoye15}. 
Quantum friction is associated with the emission and the propagation of electromagnetic waves in the medium 
at a velocity different from the speed of light in vacuum and the physics behind it shows connections with the Vavilov-Cherenkov effect \cite{Cerenkov37,Frank37,Silveirinha13,Maghrebi13a,Pieplow15}. 
The analogies between all the above-mentioned phenomena naturally generate questions about the 
similarities of the underlying physical mechanisms \cite{Frolov86,Ginzburg96}.

\begin{figure}
\center
\includegraphics[width=0.49\textwidth]{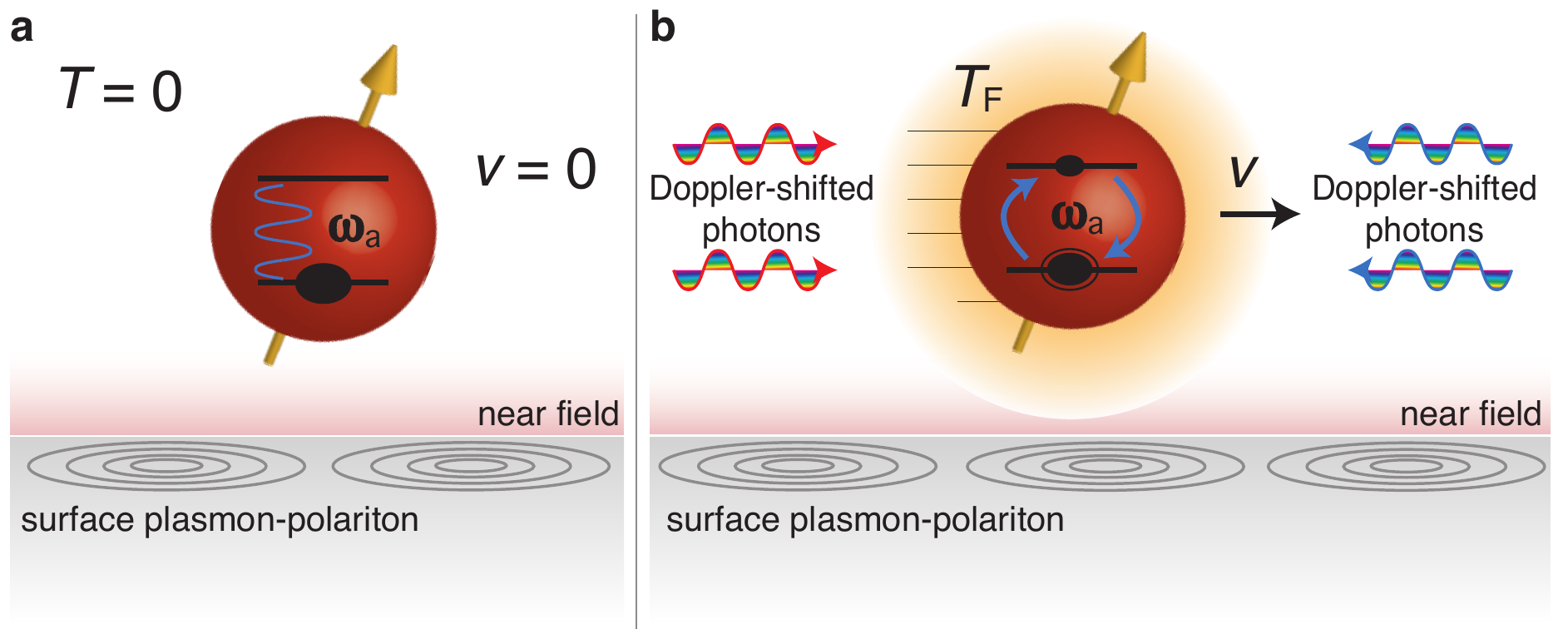}
\vspace{-.9cm}
\caption{Stationary versus moving atom. When the atom is stationary ($v=0$) the system is in its global ground state (panel a). The motion at constant velocity parallel to a surface induces a thermal-like steady-state characterised by the temperature $T_{\rm F}$ determined by the atom's velocity and distance from the surface (see Eq.~\eqref{TvReal2} (panel b).
}\label{FrictionT}
\end{figure}

In this work we investigate some of these behaviours and we draw out a connection between quantum friction and the FDU effect. We show that, although it is moving near a surface at \emph{constant velocity}, the atom feels the surrounding vacuum as a state which is \emph{thermal} in nature (see Fig. \ref{FrictionT}). This leads us to the definition of an effective temperature with characteristics analogue to $T_{\rm HU}$.
A simple dimensional analysis allows us to anticipate the form of 
this result. We only require that, as in the case of Eq. \eqref{UnruhTemp}, the effective temperature only depends on the kinematics and, specifically to our configuration, on the geometry of our system. 
We have then
\begin{equation}
T_{\rm F}\propto\frac{\hbar v}{k_{\rm B} z_{a}}~, 
\end{equation}
where $v$ is the velocity of our moving object and $z_{a}$ is its distance from the surface. 
In the remainder of the paper we provide a derivation of this simple expression and fix the corresponding prefactor. 

The approach we follow relies on results of quantum thermodynamics and in particular on the 
statistical information encoded in the fluctuation-dissipation theorem (FDT) or rather in one of 
its extensions to the non-equilibrium dynamics that characterises frictional process 
\cite{Intravaia14,Intravaia16a}. 
We recall that, in the framework of equilibrium (quantum and classical) thermodynamics, the FDT establishes a 
connection between the power spectrum tensor  $\underline{S}(\omega)$ and the linear response (susceptibility tensor) 
$\underline{\alpha}(\omega)$ of a system to an external perturbation 
\cite{Callen51}. 
In its simplest formulation the FDT states that 
\begin{equation}
\underline{S}(\omega)=\frac{\hbar}{\pi}[n(\omega,T)+1]\underline{\alpha}_{I}(\omega)\xrightarrow{T\to 0} \frac{\hbar}{\pi}\theta(\omega)\underline{\alpha}_{I}(\omega),
\label{FDT}
\end{equation}
where $n(\omega,T)=[e^{\frac{\hbar \omega}{k_{\rm B}T}}-1]^{-1}$ is the Bose-Einstein thermal distribution.
In this expression and throughout this work, we adopt the subscript ``$I$'' (``$R$'') to denote 
the imaginary (real) part of a function or tensor. 

For our analysis we consider an atom (here synonymous for a microscopic neutral system with internal degrees of freedom) that is moving in vacuum parallel to a surface at a fixed distance	
 $z=z_{a}>0$. In the following, we assume that
the half-space $z<0$ consists of a linear dispersive and 
dissipative material.
An external force is applied to the atom and, as soon as this drive is balanced by quantum friction, 
the system reaches a non-equilibrium steady state (NESS) within which the atom moves at constant velocity $\mathbf{v}$. 
The centre of mass will then be moving along the trajectory $\mathbf{r}(t)=(\mathbf{R}(t),z_{a})$, 
where for $t\to \infty$, we have $\mathbf{R}(t)\approx \mathbf{v}t$. We represent the atom's 
internal dynamics by the dipole operator $\mathbfh{d}(t)=\mathbf{d}\hat{q}(t)$, where $\mathbf{d}$ 
is the (real) dipole vector coupling constant and $\hat{q}(t)$ is a dimensionless operator whose 
dynamics is described here in terms of a harmonic oscillator with frequency $\omega_{a}$. 
To simplify our treatment, 
we will only consider non-relativistic velocities ($v/c\ll1$, $v=|\mathbf{v}|$) and neglect all magnetic 
effects \cite{Pendry97,Volokitin02,Intravaia14}. 
In the atomic rest frame, the equation of motion for the dipole with a centre of mass moving along 
$\mathbf{r}(t)$ is
\begin{equation}
\ddot{\mathbfh{d}}(t)+\omega_{a}^{2}\mathbfh{d}(t)=\frac{2\omega_{a}}{\hbar}\mathbf{d}\mathbf{d}\cdot \mathbfh{E}(\mathbf{r}(t),t) ~.
\label{eqmotion}
\end{equation}
The operator $\mathbfh{E}(\mathbf{r},t)$ describes the total electromagnetic field solution of the Maxwell equations and it is in general given by sum of different contributions: one is $\mathbfh{E}_{0}(\mathbf{r},t)$, the field generated solely by the surface and due to the (quantum) fluctuating currents in the medium \cite{Rytov53,Intravaia12b}; the other is the field induced by the dipole itself and scattered by the surface \cite{Intravaia11}.
In frequency space, the stationary solution, $\mathbfh{d}(\omega)$, of the equation of motion \eqref{eqmotion} can be written as the product of the Dopple-shifted field  $\mathbfh{E}_{0}$ and of the velocity dependent (dressed) polarisability tensor \cite{Intravaia14,Intravaia16a},
 \begin{equation}
\underline{\alpha}(\omega,\mathbf{v})=
\frac{\frac{2\omega_{a}}{\hbar(\omega_a^2 -\omega^2)}\mathbf{d}\mathbf{d}}{1 -\frac{2\omega_{a}}{\hbar(\omega_a^2 -\omega^2)} \int \frac{d^2{\bf k}}{(2 \pi)^{2}} {\bf d} \cdot  \underline{G}(\mathbf{k},z_{a},\mathbf{k}\cdot\mathbf{v}+\omega) \cdot 
{\bf d}}~.
\label{alpha}
\end{equation}
Here, $\mathbf{k}=(k_{x},k_{y})$ is the wave vector parallel to the surface and 
$\underline{G}(\mathbf{k},z_{a}; \omega)$ 
is the spatial Fourier transform of the electromagnetic Green tensor in the $xy$-plane. It contains the information about the geometry of the system and also describes the interaction between the field and the medium composing the surface. 
In the steady-state, the dipole's velocity dependent power spectrum  can be obtained from the dipole's correlation tensor 
$\langle\mathbfh{d}(\omega)\mathbfh{d}(\omega')\rangle=4\pi^{2}\underline{S}_{v}(\omega) \delta(\omega+\omega')$,
where the average is taken over the initial factorized state of the system \cite{Van-Kampen92}. 
Since the polarizability is not a quantum operator, the power spectrum $\underline{S}_{v}(\omega)$ turns to be related to the correlation tensor of $\mathbfh{E}_{0}(\mathbf{k},z_{a}; \omega)$ with respect to the initial state of field. This is obtained from the FDT, where in this case the susceptibility is given by the Green tensor.
Assuming that both subsystems are initially in their ground state, we obtain (see Refs. \cite{Intravaia14,Intravaia16a} for more detail)
\begin{multline}
\underline{S}_{v}(\omega)=\frac{\hbar}{\pi}\int \frac{d^{2}\mathbf{k}}{(2\pi)^{2}}\,\theta(\mathbf{k}\cdot\mathbf{v}+\omega)\\
\times\underline{\alpha}(\omega,\mathbf{v})\cdot
\underline{G}_{I}(\mathbf{k},z_{a},\mathbf{k}\cdot\mathbf{v}+\omega)\cdot\underline{\alpha}^{*}(\omega,\mathbf{v})~.
\label{spectrumNESS}
\end{multline}
Equation \eqref{spectrumNESS} shows that in the presence of the material interface and when the NESS is 
achieved, the power spectrum, unlike the FDT, is not simply proportional to the imaginary part of the susceptibility 
but rather, it is an involved function of the Green tensor and the dressed polarisability. For $v=0$ we recover
the expression in Eq. \eqref{FDT} using an identity that connects the polarizability in Eq. \eqref{alpha} with the Green tensor \cite{Intravaia14,Intravaia16a}
\begin{equation}
\underline{\alpha}_{I}(\omega,\mathbf{v})=\int \frac{d\mathbf{k}}{(2\pi)^{2}}\underline{\alpha}(\omega,\mathbf{v})\cdot \underline{G}_{I}(\mathbf{k},z_{a};\mathbf{k}\cdot\mathbf{v}+\omega)\cdot \underline{\alpha}^{*}(\omega,\mathbf{v}).
\label{Imalpha}
\end{equation}

Interestingly, this same identity allows to rewrite Eq. \eqref{spectrumNESS} in a form 
which is formally identical to the expression in Eq. \eqref{FDT},
\begin{equation}
\underline{S}_{v}(\omega)=\frac{\hbar}{\pi}[N_{v}(\omega)+1]\underline{\alpha}_{I}(\omega,\mathbf{v})~.
\label{Unruh}
\end{equation}
The effective occupation number is defined as $N_{v}(\omega)
=\underline{J}(\omega,\mathbf{v})\cdot\underline{\alpha}^{-1}_{I}(\omega,\mathbf{v})$, where $\underline{J}(\omega,\mathbf{v})$ is essentially the difference between the zero-temperature FDT in \eqref{FDT} and the expression in \eqref{spectrumNESS} 
\cite{Note1}.
The similarities between Eqs. \eqref{FDT} 
and \eqref{Unruh} allow us to wonder
whether the function $N_{v}(\omega)$ describes an effective Planckian (bosonic) thermal occupation number. As for the FDU effect, if the state felt by the atom were thermal, the effective temperature obtained by inverting $N_{v}(\omega)=n(\omega,T_{v})$ would be constant. After some algebra the solution of this equation can be written as
\begin{equation}
T_{v}(\omega)=\left(\frac{\hbar \omega}{k_{\rm B}}\right)/\log\left[\frac{\Sigma^{(0)}(\omega)+\Sigma^{+}_{v}(\omega)}{\Sigma^{-}_{v}(\omega)}\right]~.
\label{teff}
\end{equation}
where we have defined the functions
\begin{equation}
 \Sigma^{\pm}_{v}(\omega) = \int \frac{d^{2}\mathbf{k}}{(2\pi)^{2}}
                            \theta(\mathbf{k}\cdot\mathbf{v}\pm\omega)
                             \mathbf{d}
                            	\cdot
                            	\underline{g}_{I}(\mathbf{k},z_{a}; \mathbf{k}\cdot\mathbf{v}\pm\omega)
                            	\cdot
                            \mathbf{d}~.
 \label{SigmaDef}                           
\end{equation} 
In the previous expressions, $\underline{g}(\mathbf{k},z_{a}; \omega)$ is the scattering part of the Green tensor related 
to the field reflected by the material interface \cite{Wylie84}. The function $\Sigma^{(0)}(\omega)$ is connected to 
the atom's spontaneous decay in vacuum: 
In our point-like description of the dipole $\Sigma^{(0)}(\omega)=(\omega/c)^{3}|\mathbf{d}|^{2}/(6\pi\epsilon_{0})$ 
where $\epsilon_{0}$ is the vacuum permittivity \cite{Intravaia11a}.  
The structure of Eq. \eqref{teff} can be understood in terms of a perturbative calculation \cite{Note1}: the functions $\Sigma^{(0)}(\omega)$ and $\Sigma^{\pm}_{v}(\omega)$ are indeed related to motional-induced transition rates from ground state to the excited state, $\gamma^{-}_{v}$, and from the excited state to the ground state, $\gamma^{+}_{v}$. The temperature in Eq. \eqref{teff} results from the condition $\exp[\hbar \omega_{a}/(k_{B}T_{v})]=\gamma^{+}_{v}/\gamma^{-}_{v}$ characterizing the thermal equilibrium for the populations of the atomic levels  \cite{Note1,Breuer02}.

%
\begin{figure}
\includegraphics[width=8.7cm]{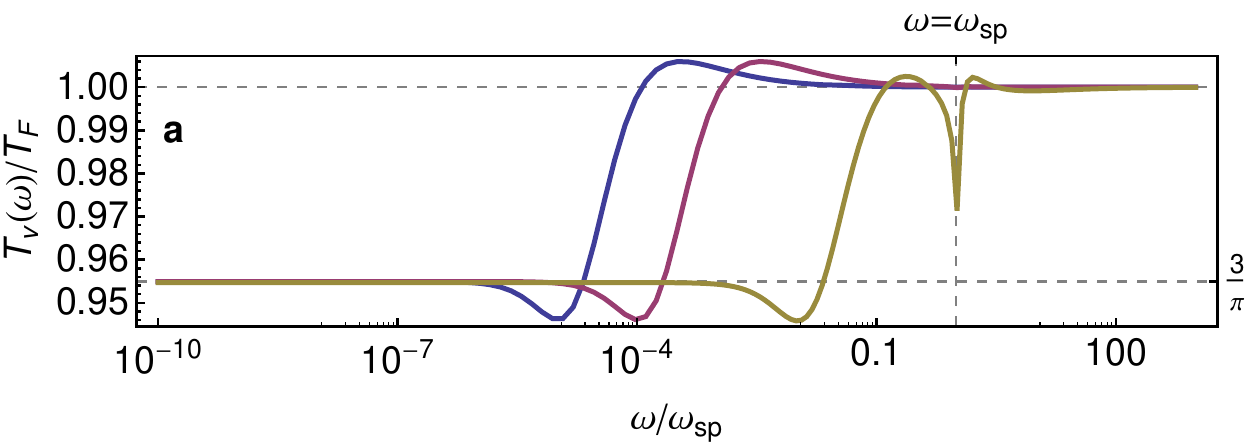}
\includegraphics[width=8.7cm]{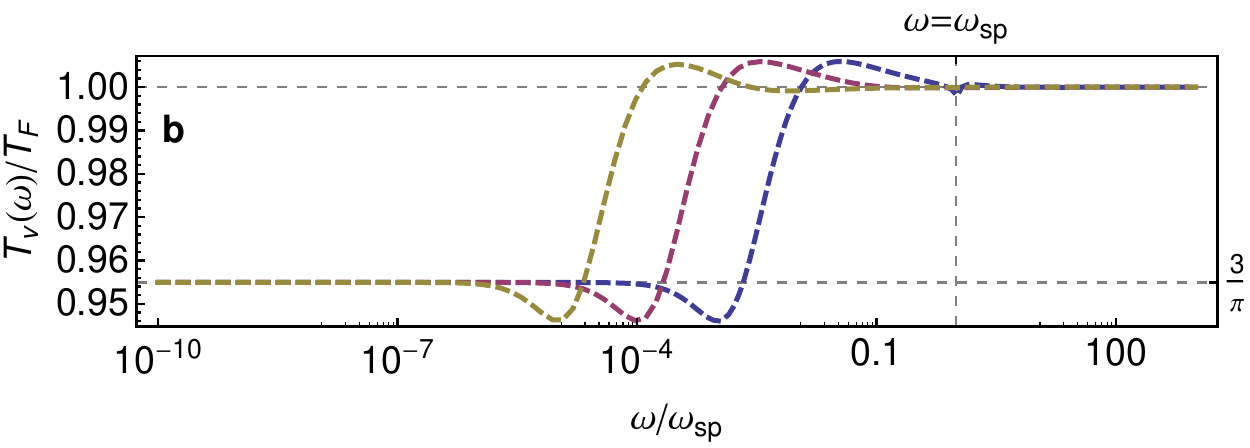}
\includegraphics[width=8.7cm]{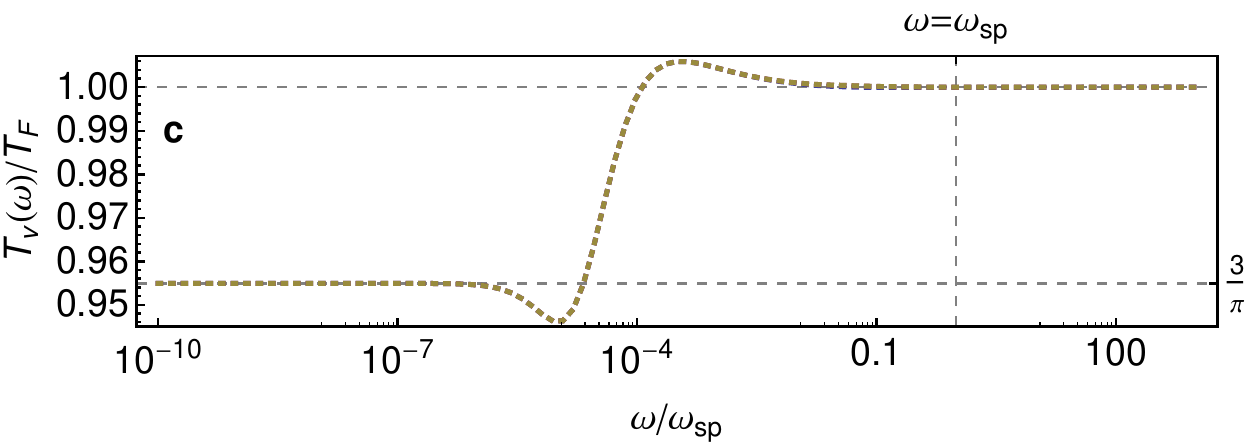}
\vspace{-.7cm}
\caption{Frequency dependence of the effective temperature. The expression in Eq. \eqref{teff} is averaged over all dipole directions and
         normalised to 
         $T_{\rm F}=\hbar v/(2 k_{\rm B} z_{a})$. The surface's material is a metal described by
         the Drude model $\epsilon(\omega)=1-\omega_{p}^{2}[\omega(\omega+\imath \Gamma)]^{-1}$
          with dissipation rate $\Gamma$ and plasma frequency $\omega_{p}$.
         The surface plasmon frequency is then $\omega_{\rm sp}=\omega_{p}/\sqrt{2}$.
         Panel a: The graph depicts results for 
         different values of the atom's velocity for a constant distance 
         $z_{a}\omega_{\rm sp}/c=10^{-1}$ and fixed dissipation in the metal 
         ($\Gamma/\omega_{\rm sp}=10^{-2}$): 
         Blue line $v/c=10^{-6}$; purple line $v/c=10^{-5}$; 
         yellow line $v/c=10^{-3}$. 
         The dip around $\omega\sim \omega_{\rm sp}$ corresponds to the excitation 
         of a surface plasmon-polariton at the vacuum/metal interface. 
         Panel b: The behaviour of the averaged effective temperature is similar for different atom-surface distances
         for the same fixed dissipation and 
         constant velocity $v/c=10^{-4}$: 
         Blue dashed line $z_{a}\omega_{\rm sp}/c=10^{-1}$; purple dashed line 
         $z_{a}\omega_{\rm sp}/c=1$; yellow dashed line $z_{a}\omega_{\rm sp}/c=10^{1}$.
         Panel c: Finally, we depict the results for different dissipations rates 
         ($\Gamma/\omega_{\rm sp}=10^{-6},10^{-3},10^{-1}$)
         of the metal for constant velocity ($v/c=10^{-6}$) and distance 
         ($z_{a}\omega_{\rm sp}/c=10^{-1}$). This graph demonstrates that the effective 
         temperature is almost insensitive to the properties of the metal.
         In all the previous plots the effective temperature show a little jump
         around $\omega\sim v/z_{a}$.
}\label{EffectiveTPlot}
\end{figure}
%

In order to obtain further insight into the dependence of the effective temperature on the system
parameters we consider the case where the atom moves within the near field at the material interface (see Fig. \ref{FrictionT}).
In this case we can approximate the scattered Green tensor by its quasi-electrostatic limit 
\cite{Joulain05,Intravaia11}.
The relevant part of the tensor can be written as (SI units)
\begin{equation}
 \underline{g}_{I}(\mathbf{k},z_{a};\omega)\approx \frac{r^{p}_{I}(\omega)}{2\epsilon_{0}}k e^{-2k z_{a}} \left(\frac{k_{x}^{2}}{k^{2}}\mathbf{x}\mathbf{x}+\frac{k_{y}^{2}}{k^{2}}\mathbf{y}\mathbf{y}+\mathbf{z}\mathbf{z}\right)
 \label{GreenNearField}
\end{equation}
Here, we have introduced $k=|\mathbf{k}|=\sqrt{k_{x}^{2}+k_{y}^{2}}$, while the unit vectors $\mathbf{x}$, $\mathbf{y}$ and $\mathbf{z}$ indicate the spatial directions. As usual, in this limit ($c\to \infty$), only the p-polarised reflection coefficient, $r^{p}(\omega)$, is relevant to the description and it only depends on the frequency \cite{Intravaia11}. 
It is interesting to extract the expressions for the low- and high-frequency limits of 
$T_{v}(\omega)$. The former value is determined by the behavior of $\underline{g}_{I}(\mathbf{k},z_{a}; \omega)$ for 
$\omega < v/z_{a}$. For realistic values, $v/z_{a}$ is a frequency belonging to the regime where most materials are ohmic. This allows us to use the approximaiton
$r^{p}_{I}(\omega) \propto \omega$, and after averaging over all dipole orientations, we obtain
\begin{equation}
 T_{v}(\omega)\xrightarrow[\text{average}]{\omega\to 0} \frac{3}{\pi} \frac{\hbar v}{2k_{\rm B} z_{a}}.
 \label{TvReal}
\end{equation}
Notice that, within our description, the above result neither depends on the properties 
of the material nor on the parameters that characterise the atom's internal dynamics. Instead,
the effective temperature is solely determined by kinematics and geometry, i.e., by the atom's 
velocity and its distance from the surface of the material. The reason for such a behaviour becomes more clear when analysed in terms of
transition rates: at low frequency $\gamma^{\pm}_{v}$ are both proportional to the 
dipole strength and the material damping which then factor out in the effective temperature.
In the opposite limit, $\omega\to \infty$, the function $\Sigma^{+}_{v}(\omega)$ vanishes
algebraically with $\omega$ and $z_{a}$. Conversely, the behaviour of the Green tensor for a surface 
as a function of the lateral wave vector (see Eq. \eqref{GreenNearField} and \cite{Wylie84,Note1}) implies that $\Sigma^{-}_{v}(\omega)$ 
vanishes exponentially as $\exp[-2\omega z_{a}/v]$. Because the vacuum contribution $\Sigma^{(0)}(\omega)$ 
only grows as a power law for $\omega \to \infty$, we have
\begin{equation}
 T_{v}(\omega)\xrightarrow{\omega\to \infty} T_{\rm F}= \frac{\hbar v}{2k_{\rm B} z_{a}}~,
 \label{TvReal2}
\end{equation}
which is only a factor $\pi/3\sim 1.04$ (4\%) larger than the value obtained in low-frequency limit
\eqref{TvReal}. Once again, the high-frequency limit only depends on the kinematics and the
geometrical properties of the system and is insensitive to the material's and the atom's degrees of freedom.
In Fig.~\ref{EffectiveTPlot}, we depict the function $T_{v}(\omega)$, averaged over all dipole's directions and normalised by its high-frequency 
limit, for the case of a dissipative metal described by the Drude model 
\cite{Jackson75}.  
We observe that $T_{v}(\omega)$ varies only by a few percent across the entire frequency range, 
thereby changing from its low-frequency value to its high-frequency limit around $\omega\sim v/z_{a}$.
Depending on the parameters, a sharp change in the effective temperature is also visible for $\omega\sim \omega_{\rm sp}$ (cfr. Fig.~\ref{EffectiveTPlot}a) 
which corresponds to the excitation of surface plasmon-polaritons at the vacuum/metal interface 
\cite{Joulain05,Intravaia05}.
Interestingly, the whole function $T_{v}(\omega)$ is rather insensitive to the value of the dissipation rate in the metal (cfr. Fig.~\ref{EffectiveTPlot}c) and one can show that it only depends on general features 
(sub-ohmic, ohmic or super-ohmic) of the
dissipative process in the medium \cite{Intravaia_inprep}.

As mentioned above, a nearly constant effective temperature denotes that $N_{v}(\omega)$ behaves 
as a thermally equilibrated bosonic occupation number indicating that, during its motion, the atom feels the surrounding vacuum as if it were a thermal state. This behaviour is 
analogous to the FDU effect \cite{Boyer84,Alsing04} with the important difference that the atom is 
moving at \emph{constant velocity} rather than with uniform acceleration. 
Also, our approach does not depend on the initial atomic trajectory which therefore does not affect the temperature in the NESS \cite{Intravaia15}.
The physical mechanism responsible for the atom's incalescence can be identified in the so-called anomalous Doppler effect  \cite{Frolov86,Ginzburg96}: Mathematically this describes a change in the sign of the Doppler shifted radiation's frequency. 
Physically, in this process (also occurring in the quantum Vavilov-Cherenkov effect) the atom's kinetic energy is converted into radiation and part of it can also increase its internal energy  \cite{Frolov86,Ginzburg96}. 
Importantly, however, we showed that the expression for $T_{\rm F}$ does not simply describe a temperature equivalent of the transferred energy but rather, in agreement with the expression for $N_{v}(\omega)$, it indicates that some form of thermalization is occurring when the stationary non-equilibrium is achieved \cite{Chetrite13,Nandkishore15,Eisert15}. 
This prompts the effect described above as an interesting and experimentally accessible case of study for investigating how quantum systems driven out of equilibrium behave in their NESSs.

The insensitivity of the final result to any specific parameter associated with the atom's 
internal degrees of freedom denotes that its validity may transcend the specific model for 
the dipole used here \cite{Note2}.
Notice also that, although it is not evident from the Eqs. \eqref{TvReal} and \eqref{TvReal2}, real material properties, such as dissipation and dispersion, play an important role in our derivation by providing a non-vanishing expression for Eq. \eqref{GreenNearField}. In order to further understand this point a surface made by an ideal material with real constant positive permittivity ($\epsilon=n_{\epsilon}^{2}>0$) can be considered. In this case a description in terms of near-field is no longer appropriate and the full expression for the Green tensor must be used. Using Eq. \eqref{teff} one can show that, even in this limit, an effective temperature can still be defined as in Eq. \eqref{TvReal2} where, however, the replacement $v\to \theta\left(v-c/n_{\epsilon}\right)\left[v-c/n_{\epsilon}\right]$ has to be made \cite{Note1}. This expression is directly associated with the Vavilov-Cherenkov radiation and it shows that the temperature is nonzero only when the corresponding velocity threshold, $v>c/n_{\epsilon}$, is met \cite{Frolov86,Ginzburg96,Silveirinha13,Maghrebi13a,Pieplow15}  (for $n_{\epsilon}\gg 1$ non-relativistic velocities can still be considered). The main impact of realistic material properties is therefore to affect the threshold imposed by the speed of light in the medium, allowing for it to be zero.

It is also important to stress the role of non-equilibrium physics in our analysis. The FDT-like 
expression in Eq.~\eqref{Unruh} reveals two major modifications relative to equilibrium physics, both
of which are connected to the motion of the atom.  
The first is the appearance of Doppler-shifted frequencies which, in turn, leads to the 
occurrence of the anomalous Doppler effect and to the excitation of the atom's internal 
degrees of freedom 
\cite{Frolov86,Ginzburg96}. 
However, Eq.~\eqref{Unruh} is not simply a Doppler-shifted version of Eq. \eqref{FDT} and the additional 
modification induced by non-equilibrium physics plays an essential role into the appearance of the effective 
temperature (see Ref. \cite{Intravaia14,Intravaia16a,Note1}). 

Finally, since the physical process that ``heats'' the atom is the same that leads to 
quantum friction, measuring the former leads to an indirect investigation of the latter. 
For an atom moving at a velocity of 340 m/s at a distance of 10 nm from the surface, the 
effective temperature $T_{\rm F}$ is about 130 mK,  
which in frequency units corresponds to $\omega_{\rm F}/(2\pi)\approx 2.7$ GHz. 
Therefore, measurements of the internal state population in systems involving, for example, beams of atoms \cite{Sandoghdar92}, molecules \cite{Arndt14,Cotter15} or even defect-centres in nano-diamonds \cite{Jelezko06,Schell14} can be suitable candidates for the experimental investigation of the previous predictions. 

\paragraph*{Acknowledgments.}
We are indebted with K. Busch, R. O. Behunin and D. A. R. Dalvit for their support and for providing many important comments and suggestions during the preparation of this work.
We are also thankful to B.-L. Hu and R. Onofrio for reading the manuscript and for useful discussions which led to several improvements. 
We acknowledge financial support from the European Union Marie Curie People program through the Career Integration Grant No. PCIG14- GA-2013-631571 and from the DFG through the DIP program (FO 703/2-1).



\newpage
\cleardoublepage
\setcounter{page}{1}
\setcounter{figure}{0}
\setcounter{equation}{0}
\renewcommand{\theequation}{S\arabic{equation}}
\renewcommand{\figurename}{\textbf{Supplementary Figure}}
\renewcommand{\thefigure}{{\bf S\arabic{figure}}}

\section*{Supplemental Material}

\subsection{Power spectrum and effective temperature}

In Refs. \cite{Intravaia14,Intravaia16a} it was showed that, for an oscillator moving with constant velocity above a surface, the dipole power spectrum is given by Eq. \eqref{spectrumNESS}. Using the identity in Eq. \eqref{Imalpha} allows us to recast $\underline{S}_{v}(\omega)$ as
\begin{equation}
\underline{S}_{v}(\omega)=\frac{\hbar}{\pi}\left[\theta(\omega)\underline{\alpha}_{I}(\omega,\mathbf{v})+\underline{J}(\omega,\mathbf{v})\right]~,
\label{S-NEFDT}
\end{equation}
where we have introduced
\begin{multline}
\underline{J}(\omega,\mathbf{v})=\int \frac{d^{2}\mathbf{k}}{(2\pi)^{2}}\left[\theta(\mathbf{k}\cdot\mathbf{v}+\omega)-\theta(\omega)\right]\\
\times\underline{\alpha}(\omega,\mathbf{v})\cdot
\underline{G}_{I}(\mathbf{k},z_{a},\mathbf{k}\cdot\mathbf{v}+\omega)\cdot\underline{\alpha}^{*}(\omega,\mathbf{v})~.
\end{multline}
If we define the effective occupation number $N_{v}(\omega)
=\underline{J}(\omega,\mathbf{v})\cdot\underline{\alpha}^{-1}_{I}(\omega,\mathbf{v})$,
equation \eqref{S-NEFDT} takes the form FDT-like form given in Eq. \eqref{Unruh}.

The Green tensor can be written as 
$\underline{G}(\mathbf{k},z_{a}; \omega)=\underline{G}^{(0)}(\mathbf{k}; \omega)+\underline{g}(\mathbf{k},z_{a}; \omega)$, 
i.e., as the sum of the vacuum contribution ($\underline{G}^{(0)}$) and 
of interface-induced scattered part ($\underline{g}$).
Despite our non relativistic treatment, we have to preserve the vacuum's Lorentz invariance. 
Practically, this requires that all contributions containing $\underline{G}^{(0)}$ must behave 
similar to their static ($v=0$) counterpart. For example, we have ($\omega>0$)
\begin{multline}
 \int \frac{d^{2}\mathbf{k}}{(2\pi)^{2}}\theta(\mathbf{k}\cdot\mathbf{v}+\omega)\mathbf{d}\cdot\underline{G}_{I}^{(0)}(\mathbf{k},z_{a}; \mathbf{k}\cdot\mathbf{v}+\omega)\cdot\mathbf{d}\\
\to \int \frac{d^{2}\mathbf{k}}{(2\pi)^{2}}\mathbf{d}\cdot\underline{G}_{I}^{(0)}(\mathbf{k},z_{a};\omega)\cdot\mathbf{d}
=\frac{|\mathbf{d}|^{2}}{6\pi\epsilon_{0}}\left(\frac{\omega}{c}\right)^{3}\equiv \Sigma^{(0)}(\omega),
\label{sigmanull}
\end{multline}
and similarly for the case without the Heaviside function. Physically this means that, 
neglecting terms in $v/c$, the empty-space 
spontaneous decay is not affected by the motion of the particle. 
Using that $\theta(\mathbf{k}\cdot\mathbf{v}+\omega)-1=-\theta(-\mathbf{k}\cdot\mathbf{v}-\omega)$ for $\omega>0$, 
we can write
\begin{align}
\underline{J}(\omega,v)&=-\int \frac{d^{2}\mathbf{k}}{(2\pi)^{2}}\theta(-\mathbf{k}\cdot\mathbf{v}-\omega)
\nonumber\\
&\times
\underline{\alpha}(\omega,v)\cdot
\underline{g}^{\rm s}_{I}(\mathbf{k},z_{a},\mathbf{k}\cdot\mathbf{v}+\omega)\cdot\underline{\alpha}^{*}(\omega,v)\nonumber\\
&=\int \frac{d^{2}\mathbf{k}}{(2\pi)^{2}}\theta(\mathbf{k}\cdot\mathbf{v}-\omega)
\nonumber\\
&\times
\underline{\alpha}(\omega,v)\cdot
\underline{g}^{\rm s}_{I}(\mathbf{k},z_{a},\mathbf{k}\cdot\mathbf{v}-\omega)\cdot\underline{\alpha}^{*}(\omega,v)~.
\end{align}
In the previous expression the vacuum term drops out for the argument after Eq. \eqref{sigmanull}. We also used that 
$\underline{g}^{\rm s}_{I}(-\mathbf{k},z_{a},\omega)=\underline{g}^{\rm s}_{I}(\mathbf{k},z_{a},\omega)$ and 
$\underline{g}^{\rm s}_{I}(\mathbf{k},z_{a},-\omega)=-\underline{g}^{\rm s}_{I}(\mathbf{k},z_{a},\omega)$, 
which reflect the symmetry properties of the Green tensor for a flat material interface. (The superscript ``${\rm s}$'' was introduced to emphasize that only the symmetric part of the tensor is relevant here \cite{Intravaia14,Intravaia16a}.) 
Using the same symmetry properties of the Green tensor and of the atom's polarisability one can also show that, in full analogy with the bosonic occupation number, $N_{v}(\omega)$ satisfies the identity ($\omega>0$)
\begin{equation}
 -N_{v}(-\omega)=N_{v}(\omega)+1~.
\end{equation}
which, in agreement with the usual FDT, extends Eq. \eqref{Unruh} to negative frequencies.
The definitions in Eq. \eqref{SigmaDef} also
allow us to write the effective bosonic thermal occupation number as
\begin{align}
N_{v}(\omega)
=\frac{1}{\left[\Sigma^{(0)}(\omega)+\Sigma^{+}_{v}(\omega)\right]/\Sigma^{-}_{v}(\omega)-1}~.
\label{S-MotionN}
\end{align}
From a direct comparison of the previous equation with the expression for $n(\omega,T)=[e^{\hbar \beta \omega}-1]^{-1}$,
we infer the definition for the effective temperature given in Eq. \eqref{teff}.

At non-zero velocity, a Taylor expansion up to the first order gives
$ \Sigma^{\pm}_{v}(\omega)\approx\Sigma_{v}(0)\pm\sigma_{v}\omega$,
where 
$\Sigma_{v}(0)=\Sigma^{+}_{v}(0)=\Sigma^{-}_{v}(0)$ and $\sigma_{v}=\partial_{\omega}\Sigma^{+}_{v}(\omega)\vert_{\omega=0}$. 
For $\omega\to 0$ we have $\Sigma^{(0)}(\omega)\propto \omega^{3}$ and we thus obtain
\begin{equation}
T_{v}(\omega\to 0)=
\frac{\hbar }{2k_{\rm B}}\frac{\Sigma_{v}(0)} {\sigma_{v}}.
\end{equation}
Using Eq. \eqref{GreenNearField} with $r^{p}_{I}(\omega) \propto \omega$, and averaging over the dipole's direction, we obtain the expression in Eq. \eqref{TvReal}.

For $\omega\to \infty$, in the near-field limit, we have
\begin{align}
  \Sigma_{v}^{-}(\omega) \stackrel{\omega\to \infty}{\sim} P(\omega) \left(\frac{\omega z_{a}}{v}\right)^{\frac{3}{2}}e^{-2\frac{\omega z_{a}}{v}}~,
\label{lageSigmaM}
\end{align} 
where $P(\omega)$ is a polynomial function of the frequency. The detail of this function is, however, irrelevant for determining the high frequency limit of Eq.\eqref{teff}, which is rather determined by the exponential function in Eq. \eqref{lageSigmaM}.  
Within the same limit and approximation we have 
$\Sigma_{v}^{+}(\omega) \stackrel{\omega\to \infty}{\sim}r^{p}_{I}(\omega)$, which indicates that this function goes to zero at high freuquency.
If we now insert Eq. \eqref{lageSigmaM} in Eq.\eqref{teff}, using that $\Sigma^{(0)}(\omega)$ diverges as a power law of $\omega$, in the limit $\omega
\to \infty$ we recover the expression given in Eq.\eqref{TvReal2}.

\subsection{Perturbative analysis}

Equation \eqref{teff} can also be derived within a time-dependent perturbative approach \cite{Intravaia15}. We sketch here the main steps that lead to the effective temperature and compare the result with that obtained in the main text. Unlike the main text, here we model the atom's internal degrees of freedom in terms of a two-state system. 

The Hamiltonian of our system is characterised by the time-dependent interaction term
\begin{equation}
\hat{V}(t)=-\mathbfh{d}\cdot\mathbfh{E}(\mathbf{r}(t)).
\end{equation}
We will be working in the Schr\"odinger picture. 
In time-dependent perturbation theory if $c_{i \to f}(t)$ describes the transition amplitude for the transition $|i\rangle\to |f\rangle$ the corresponding transition rate is defined as 
\begin{equation}
\gamma_{i\to f}=\lim_{t\to \infty}\frac{|c_{i \to f}(t)|^{2}}{t}~.
\end{equation}

We are interested in the rates for the transitions $|g,\{0\}\rangle \to |e,\{n\}\rangle$ and $|e,\{0\}\rangle \to|g,\{n\}\rangle$, where $|\{n\}\rangle$ is a generic state of the field. 
The total transition rate is obtained by summing over all states of the field (only those energetically compatible are selected in the limit $t\to \infty$). For simplicity of notation, we will indicate as $\gamma^{-}_{v}$ the total transition rate to the excited state and $\gamma^{+}_{v}$ the total transition rate to the ground state. From the theory of open quantum systems we can relate these transition rates to the effective temperature through the relation $\gamma^{+}_{v}/\gamma^{-}_{v}=\exp[\hbar \omega_{a}/(k_{B}T_{v})]$. This leads to the definition
\begin{equation}
T_{v}=\frac{\hbar \omega_{a}}{k_{B}}/\log[\frac{\gamma^{+}_{v}}{\gamma^{-}_{v}}]~,
\label{S-tempPert}
\end{equation}
where $\omega_{a}$ is the particle's internal transition frequency.

Within our model, the dipole operator is written as $\mathbfh{d}=\mathbf{d}\hat{\sigma}_{1}$ where $\hat{\sigma}_{1}$ is the Pauli matrix connecting the ground and the excited state. 
Using the previous expressions we have then
\begin{widetext}
\begin{multline}
|c_{\pm}(t)|^{2}= \frac{1}{\hbar^{2}}\sum_{\{n\}}\int_{t_{\rm in}}^{t} dt_{1}\int_{t_{\rm in}}^{t} dt_{2}
\int_{-\infty}^{\infty}\frac{d\omega}{2\pi}\int \frac{d^{2}\mathbf{k}}{(2\pi)^{2}} \int_{-\infty}^{\infty}\frac{d\omega'}{2\pi}\int \frac{d^{2}\mathbf{k}'}{(2\pi)^{2}}\\
\mathbf{d}\cdot  \langle \{0\}|\mathbfh{E}^{\dag}_{0}(\mathbf{k},z_{a}; \omega)|\{n\}\rangle  \langle \{n\}|\mathbfh{E}_{0}(\mathbf{k}',z_{a}; \omega')|\{0\}\rangle \cdot\mathbf{d} \\
e^{\pm\imath \omega_{a}t_{2}} e^{-\imath[\mathbf{k}\cdot\mathbf{R}(t_{2})-\omega t_{2}]}e^{\mp\imath \omega_{a}t_{1}} e^{\imath[\mathbf{k}'\cdot\mathbf{R}(t_{1})-\omega' t_{1}]}
\end{multline}
\end{widetext}
If we sum over all photonic states $|\{n\}\rangle$, we can use the FDT for evaluating the correlation function of the field
over its ground state \cite{Intravaia14,Intravaia16a} 
\begin{multline}
\langle \{0\}| \mathbfh{E}^{\dag}_{0}(\mathbf{k},z_{a},\omega)\mathbfh{E}_{0}(\mathbf{k}',z_{a},\omega')|\{0\}\rangle\\
=2(2\pi)^{3}\hbar\theta(-\omega)\underline{G}_{\Im}(-\mathbf{k},z_{a},-\omega)\delta(\omega-\omega')\delta(\mathbf{k}-\mathbf{k}')~,
\label{S-FDTfield}
\end{multline}
where we have defined the tensor where we have defined the tensor 
\begin{align}
\underline{G}_{\Im}(\mathbf{k},z;\omega)&=\frac{\underline{G}(\mathbf{k},z;\omega)-\underline{G}^{\dag}(\mathbf{k},z;\omega)}{2\imath}~.
\label{S-GImSplit}
\end{align}
Using
that $\mathbf{d}\cdot  \underline{G}_{\Im}(\mathbf{k},z_{a},\omega)\cdot\mathbf{d}=\mathbf{d}\cdot\underline{G}_{I}(\mathbf{k},z_{a},\omega)\cdot\mathbf{d}=\mathbf{d}\cdot\underline{G}_{I}(-\mathbf{k},z_{a},\omega)\cdot\mathbf{d}$ and taking the limit $-t_{\rm in},t\to \infty$, we obtain
\begin{equation}
 \gamma^{\pm}_{v}=\frac{2(2\pi)^{3}}{\hbar}
\int \frac{d^{2}\mathbf{k}}{(2\pi)^{2}}
\theta(\mathbf{k}\cdot\mathbf{v}\pm\omega_{a})\mathbf{d}\cdot\underline{G}_{I}(\mathbf{k},z_{a},\mathbf{k}\cdot\mathbf{v}\pm\omega_{a}) \cdot\mathbf{d}~.
\end{equation}
By comparing the previous expressions with the definitions in Eq. \eqref{SigmaDef} of the paper, we have
\begin{gather}
 \gamma^{+}_{v}=\frac{2(2\pi)^{3}}{\hbar}\left[\Sigma_{0}(\omega_{a})+\Sigma^{+}_{v}(\omega_{a})\right],\quad
  \gamma^{-}_{v}=\frac{2(2\pi)^{3}}{\hbar}\Sigma^{-}_{v}(\omega_{a})
\end{gather}
showing that equation \eqref{S-tempPert} exactly reproduces the effective temperature defined in equation \eqref{teff}.

Unlike the calculation described  in the main text, in the approach presented here the frequency is fixed by the internal resonance. This is a consequence of the perturbative scheme. Indeed, within our perturbative approach we have to use the bare polarizability, 
\begin{equation}
\underline{\alpha}^{(0)}(\omega,\mathbf{v})=\frac{\frac{2\omega_{a}}{\hbar}\mathbf{d}\mathbf{d}}{\omega_a^2 -(\omega+\imath 0^{+})^2}
\end{equation}
which leads to
\begin{equation}
\underline{\alpha}_{I}^{(0)}(\omega,\mathbf{v})=\frac{\mathbf{d}\mathbf{d}}{\hbar}\pi\left[\delta(\omega_{a}-\omega)-\delta(\omega_{a}+\omega)\right]~.
\end{equation}
If we consider for simplicity only positive frequencies we have then
\begin{align}
\underline{S}_{v}(\omega)\approx\frac{\hbar}{\pi}[N_{v}(\omega_{a})+1]\underline{\alpha}^{(0)}_{I}(\omega,\mathbf{v})~,
\end{align}
i.e. the previous expression is included in equation \eqref{Unruh}. For negative frequencies the derivation proceed in a similar way. 

\subsection{Vavilov-Cherenkov limit}

The previous analysis for the effective temperature can also be applied to the case of a surface made by a
medium described by a large constant real dielectric function $\epsilon=n^{2}_{\epsilon}$. 
In this case the near-field approximation is inadequate and one
has to consider the full expression for for the scattered Green tensor \cite{Wylie84,Intravaia16}

\begin{equation}
\underline{g}(\mathbf{k},z_{a}; \omega)=
\frac{\kappa}{2\epsilon_{0}}
\left[r^{p}(\omega,k)\mathbf{p}_{+}\mathbf{p}_{-}+ \frac{\omega^{2}}{c^{2}\kappa^{2}}r^{s}(\omega,k)\mathbf{s}\mathbf{s}\right]
e^{-2\kappa z_{a}}~,
\label{greenScattFull}
\end{equation}
where $\kappa^{2}=k^{2}-\omega^{2}/c^{2}$ and 
\begin{equation}
\mathbf{s}=\frac{\mathbf{k}}{k}\times \mathbf{z},\quad\mathbf{p}_{\pm}=\frac{k}{\kappa}\mathbf{z}\mp\imath\frac{\mathbf{k}}{k}~.
\end{equation}
The reflection coefficients are given by 
\begin{gather}
r^{p}(\omega,k)=\frac{n_{\epsilon}^{2}\kappa-\kappa_{m}}{n_{\epsilon}^{2}\kappa+\kappa_{m}},
\quad
r^{s}(\omega,k)=\frac{\kappa-\kappa_{m}}{\kappa+\kappa_{m}}
\label{FresnelCoeff}
\end{gather}
where $\kappa_{m}^{2}=k^{2}-n_{\epsilon}^{2}\omega^{2}/c^{2}$. The square root is chose such that $\mathrm{Re}[\kappa],\mathrm{Re}[\kappa_{m}]\ge 0$ and $\mathrm{Im}[\kappa],\mathrm{Im}[\kappa_{m}]\le 0$. Since the dyadics $\mathbf{p}_{+}\mathbf{p}_{-}$ and $\mathbf{s}\mathbf{s}$ are hermitian their symmetric part (the only relevant for our calculation) is real. 

Let us consider the function $\Sigma_{v}^{-}(\omega)$. 
Since in this case $\kappa>0$, by analysing the roots of the equation $k^{2}-n_{\epsilon}^{2}(\mathbf{k}\cdot \mathbf{v}-\omega)^{2}/c^{2}=0$, it turns out that for having non real reflection coefficients and therefore
$\Sigma_{v}^{-}(\omega)\not=0$ we must have
\begin{equation}
k>\frac{\omega}{v\cos[\varphi]-c/n_{\epsilon}}~,
\end{equation}
with $v\cos[\varphi]>c/n_{\epsilon}$, where $\varphi$ is the angle between the wave vector and the velocity. This means that $T_{v}(\omega)\not=0$ only when the Cherenkov condition is fulfilled. Indeed, a vanishing $\Sigma^{-}_{v}(\omega)$ always implies $T_{v}(\omega)=0$.

As above, it is interesting to consider some asymptotic limits of $T_{v}(\omega)$. For simplicity we will only consider large frequencies and the case $n_{\epsilon}\gg1$ so that the condition $v n_{\epsilon}/c\gtrsim 1$ is met also for non relativistic velocities.  By setting $x=c k/\omega$ we can write
\begin{multline}
 \Sigma_{v}^{-}(\omega)=\Sigma^{(0)}(\omega)\int_{-\varphi_{0}}^{\varphi_{0}}d\varphi \int_{x_{\rm M}(\varphi)}^{\infty} dx \,x^{2} f(x,v,\varphi,\mathbf{d})\\
\times  \exp\left[-2z_{a}\frac{\omega}{c}\sqrt{x^{2}-\left(x\frac{v}{c}\cos[\varphi]-1\right)^{2}}\right]
 \end{multline} 
where $\cos[\varphi_{0}]=c/(vn_{\epsilon})\lesssim 1$ (i.e. $\varphi_{0}\sim 0$) and we defined $x_{\rm M}(\varphi)=n_{\epsilon}/[(vn_{\epsilon}/c)\cos[\varphi]-1]$. The function $f(x,v,\varphi,\mathbf{d})$ has a complicated expression deriving from the definition of the scattered Green tensor given in equation \eqref{greenScattFull}. However, the only relevant aspect is that, for $x\gg 1$ it does not depend on $x$, i.e., $f(x,v,\varphi,\mathbf{d})\approx \tilde{f}(v,\varphi,\mathbf{d})$. Since $x_{\rm M}\gg 1$, for $\omega >  \left(v-c/n_{\epsilon}\right)/z_{a}$, we can write
\begin{multline}
 \Sigma^{-}_{v}(\omega)
\approx
\left[\int_{-\varphi_{0}}^{\varphi_{0}}d\varphi \,\tilde{f}(v,\varphi,\mathbf{d})  P_{3}\left(\frac{c}{z_{a}\omega}\right)\right]\\
\times\Sigma^{(0)}(\omega) \exp[-\frac{2z_{a}\omega}{v-c/n_{\epsilon}}] 
 \end{multline} 
where $P_{3}(w)$ is a third order polynomial.
The function $1+\Sigma^{+}_{v}(\omega)/\Sigma^{(0)}(\omega)$ is limited at large frequencies and therefore from equation \eqref{teff} we get that in this case
 \begin{equation}
T_{v}(\omega)\stackrel{\omega\to \infty}{\approx}\frac{\hbar}{2k_{\rm B}z_{a}}
 \theta\left(v-\frac{c}{n_{\epsilon}}\right)\left[v-\frac{c}{n_{\epsilon}}\right],
\label{CerenkovTemp}
\end{equation}
where the Heaviside function takes into account that the Cherenkov condition must be fulfilled in order to have a nonzero temperature. This is the expression described at the end of the main text. By comparing it with the expression in equation \eqref{TvReal2} it appears that real material properties (dispersion and dissipation) modify the velocity threshold allowing for it to be zero instead of $c/n_{\epsilon}$.\\

%
%


\begin{thebibliography}{10}

\bibitem{Casimir48}
H.~B.~G. Casimir, Proc. K. Ned. Akad. Wet. {\bf 51},  793  (1948).

\bibitem{Decca07}
R.~S. Decca, D. L\'{o}pez, E. Fischbach, G.~L. Klimchitskaya, D.~E. Krause, and
  V.~M. Mostepanenko, Phys. Rev. D {\bf 75},  077101  (2007).

\bibitem{Wilson11a}
C.~M. Wilson {\it et~al.}, Nature {\bf 479},  376  (2011).

\bibitem{Steinhauer14}
J. Steinhauer, Nature Phys. {\bf 10},  864  (2014).

\bibitem{Intravaia13}
F. Intravaia {\it et~al.}, Nature Commun. {\bf 4},  2515  (2013).

\bibitem{Zou13}
J. Zou {\it et~al.}, Nature Commun. {\bf 4},  1845  (2013).

\bibitem{Fulling73}
S.~A. Fulling, Phys. Rev. D {\bf 7},  2850  (1973).

\bibitem{Davies75}
P.~C.~W. Davies, J. Phys. A: Math. Gen. {\bf 8},  609  (1975).

\bibitem{Unruh76}
W.~G. Unruh, Phys. Rev. D {\bf 14},  870  (1976).

\bibitem{Hawking74}
S.~W. Hawking, Nature {\bf 248},  30  (1974).

\bibitem{Intravaia12a}
F. Intravaia, P.~S. Davids, R.~S. Decca, V.~A. Aksyuk, D. L\'opez, and D.~A.~R.
  Dalvit, Phys. Rev. A {\bf 86},  042101  (2012).

\bibitem{Crispino08}
L.~C.~B. Crispino, A. Higuchi, and G.~E.~A. Matsas, Rev. Mod. Phys. {\bf 80},
  787  (2008).

\bibitem{Boyer84}
T.~H. Boyer, Phys. Rev. D {\bf 29},  1089  (1984).

\bibitem{Alsing04}
P.~M. Alsing and P.~W. Milonni, Am. J. Phys. {\bf 72},  1524  (2004).

\bibitem{Mkrtchian03}
V. Mkrtchian, V.~A. Parsegian, R. Podgornik, and W.~M. Saslow, Phys. Rev. Lett.
  {\bf 91},  220801  (2003).

\bibitem{Dedkov05}
G. Dedkov and A. Kyasov, Phys. Lett. A {\bf 339},  212   (2005).

\bibitem{Volokitin15a}
A.~I. Volokitin, Phys. Rev. A {\bf 91},  032505  (2015).

\bibitem{Pendry97}
J.~B. Pendry, J. Phys.: Condes. Matter {\bf 9},  10301  (1997).

\bibitem{Volokitin02}
A.~I. Volokitin and B.~N.~J. Persson, Phys. Rev. B {\bf 65},  115419  (2002).

\bibitem{Dedkov02a}
G. Dedkov and A. Kyasov, Phys. Solid State {\bf 44},  1809  (2002).

\bibitem{Scheel09}
S. Scheel and S.~Y. Buhmann, Phys. Rev. A {\bf 80},  042902  (2009).

\bibitem{Barton10b}
G. Barton, New J. Phys. {\bf 12},  113045  (2010).

\bibitem{Zhao12}
R. Zhao, A. Manjavacas, F.~J. Garc{\'\i}a~de Abajo, and J.~B. Pendry, Phys.
  Rev. Lett. {\bf 109},  123604  (2012).

\bibitem{Pieplow13}
G. Pieplow and C. Henkel, New J. Phys. {\bf 15},  023027  (2013).

\bibitem{Intravaia14}
F. Intravaia, R.~O. Behunin, and D.~A.~R. Dalvit, Phys. Rev. A {\bf 89},
  050101(R)  (2014).

\bibitem{Intravaia15}
F. Intravaia, V.~E. Mkrtchian, S.~Y. Buhmann, S. Scheel, D.~A.~R. Dalvit, and
  C. Henkel, J. Phys.: Condes. Matter {\bf 27},  214020  (2015).

\bibitem{Hoye15}
J.~S. H{\o}ye, I. Brevik, and K.~A. Milton, J. Phys. A: Math. Theor. {\bf 48},
  365004  (2015).

\bibitem{Cerenkov37}
P.~A. {\v C}erenkov, Phys. Rev. {\bf 52},  378  (1937).

\bibitem{Frank37}
I. Frank and I. Tamm, C.R. Acad. Sci. URSS {\bf 14},  109  (1937).

\bibitem{Silveirinha13}
M.~G. Silveirinha, Phys. Rev. A {\bf 88},  043846  (2013).

\bibitem{Maghrebi13a}
M.~F. Maghrebi, R. Golestanian, and M. Kardar, Phys. Rev. A {\bf 88},  042509
  (2013).

\bibitem{Pieplow15}
G. Pieplow and C. Henkel, J. Phys.: Condes. Matter {\bf 27},  214001  (2015).

\bibitem{Frolov86}
V. Frolov and V. Ginzburg, Phys. Lett. A {\bf 116},  423   (1986).

\bibitem{Ginzburg96}
V.~L. Ginzburg, Physics-Uspekhi {\bf 39},  973  (1996).

\bibitem{Intravaia16a}
F. Intravaia, R.~O. Behunin, C. Henkel, K. Busch, and D.~A.~R. Dalvit, eprint:
  arXiv:1604.06405.

\bibitem{Callen51}
H.~B. Callen and T.~A. Welton, Phys. Rev. {\bf 83},  34  (1951).

\bibitem{Rytov53}
S. Rytov, {\em Theory of Electrical Fluctuations and Thermal Radiation}
  (Academy of Sciences, USSR, Moscow, 1953).

\bibitem{Intravaia12b}
F. Intravaia and R. Behunin, Phys. Rev. A {\bf 86},  062517  (2012).

\bibitem{Intravaia11}
F. Intravaia, C. Henkel, and M. Antezza,  in {\em Casimir Physics}, Vol.~834 of
  {\em Lecture Notes in Physics}, edited by D. Dalvit, P. Milonni, D. Roberts,
  and F. da~Rosa (Springer, Berlin / Heidelberg, 2011), pp.\ 345--391.

\bibitem{Van-Kampen92}
N.~G. Van~Kampen, {\em Stochastic processes in physics and chemistry}, third
  edition ed. (Elsevier, Amsterdam, 1992), Vol.~1.

\bibitem{Note1}
See Supplemental Material at [URL will be inserted by publisher] for more
  details.

\bibitem{Wylie84}
J.~M. Wylie and J.~E. Sipe, Phys. Rev. A {\bf 30},  1185  (1984).

\bibitem{Intravaia11a}
F. Intravaia, R. Behunin, P.~W. Milonni, G.~W. Ford, and R.~F. O'Connell, Phys.
  Rev. A {\bf 84},  035801  (2011).

\bibitem{Breuer02}
H. Breuer and F. Petruccione, {\em The Theory of Open Quantum Systems} (Oxford
  University Press, Oxford, 2002).

\bibitem{Joulain05}
K. Joulain, J.-P. Mulet, F. Marquier, R. Carminati, and J.-J. Greffet, Surface
  Science Reports {\bf 57},  59   (2005).

\bibitem{Jackson75}
J. Jackson, {\em Classical Electrodynamics} (John Wiley and Sons Inc., New
  York, 1975).

\bibitem{Intravaia05}
F. Intravaia and A. Lambrecht, Phys. Rev. Lett. {\bf 94},  110404  (2005).

\bibitem{Intravaia_inprep}
F. Intravaia, in preparation. 

\bibitem{Chetrite13}
R. Chetrite and H. Touchette, Phys. Rev. Lett. {\bf 111},  120601  (2013).

\bibitem{Nandkishore15}
R. Nandkishore and D.~A. Huse, Annu. Rev. Condens. Matter Phys. {\bf 6},  15
  (2015).

\bibitem{Eisert15}
J. Eisert, M. Friesdorf, and C. Gogolin, Nature Phys. {\bf 11},  124  (2015).

\bibitem{Note2}
In the Supplemental Material \cite {Note1} we show that, using a time-dependent
  perturbative calculation, we recover the same definition for the effective
  temperature in the case of an atom described in terms of a two-state system.

\bibitem{Sandoghdar92}
V. Sandoghdar, C.~I. Sukenik, E.~A. Hinds, and S. Haroche, Phys. Rev. Lett.
  {\bf 68},  3432  (1992).

\bibitem{Arndt14}
M. Arndt and K. Hornberger, Nature Phys. {\bf 10},  271  (2014).

\bibitem{Cotter15}
J.~P. Cotter {\it et~al.}, Nature Commun. {\bf 6},    (2015).

\bibitem{Jelezko06}
F. Jelezko and J. Wrachtrup, Phys. Stat. Sol. (a) {\bf 203},  3207  (2006).

\bibitem{Schell14}
A.~W. Schell, P. Engel, J.~F.~M. Werra, C. Wolff, K. Busch, and O. Benson, Nano
  Lett. {\bf 14},  2623  (2014).

\bibitem{Intravaia16}
F. Intravaia, R.~O. Behunin, C. Henkel, K. Busch, and D.~A.~R. Dalvit, eprint:
  arXiv:1603.05165.

\end{thebibliography}

\end{document}